\documentclass[aps,prb,letterpaper,showpacs,twocolumn]{revtex4}

\usepackage{graphicx}
\usepackage{latexsym}
\newcommand{\fig}[2]{\includegraphics[width=#1]{#2}}
\makeatletter
\usepackage{dcolumn}
\usepackage{bm}
\usepackage{hyperref}
\usepackage[usenames]{color}
\makeatletter \def\@dotsep{5} \makeatother
\makeatother
\usepackage{amsmath,amssymb}

\newcommand{\up}{\uparrow}
\newcommand{\dn}{\downarrow}

\begin{document}

\title{Mott and Wigner-Mott transitions in doped correlated electron systems:
effects of superlattice potential and inter-site correlation}

\author{Chunhua Li and Ziqiang Wang}

\affiliation{Department of Physics, Boston College, Chestnut Hill,
Massachusetts 02467}

\begin{abstract}

We introduce the notion of superstructure Mottness to describe the
Mott and Wigner-Mott transition in doped strongly correlated
electron systems at commensurate filling fractions away from one
electron per site. We show that superstructure Mottness emerges in
an inhomogeneous electron system when the superstructure contains an
odd number of electrons per supercell. We argue that
superstructure Mottness exists even in the absence of translation
symmetry breaking by a superlattice, provided that the extended or
intersite Coulomb interaction is strong. In the latter case,
superstructure Mottness offers a unifying framework for the Mott and
Wigner physics and a nonperturbative, strong coupling description of
the Wigner-Mott transition. We support our proposal by studying a
minimal single-band ionic Hubbard $t$-$U$-$V$-$\Delta$ model with
nearest neighbor Coulomb repulsion $V$ and a two-sublattice ionic
potential $\Delta$. The model is mapped onto a Hubbard model with
two effective ``orbitals'' representing the two sites within the
supercell, the intra and interorbital Coulomb repulsion $U$ and
$U^\prime \sim V$, and a crystal field splitting $\Delta$. Charge
order on the original lattice corresponds to orbital order.
Developing a cluster Gutzwiller approximation, we study the effects
and the interplay between $V$ and $\Delta$ on the Mott and
Wigner-Mott transitions at quarter-filling. We provide the mechanism
by which the superlattice potential enhances the correlation effects
and the tendency towards local moment formation, construct and
elucidate the phase diagram in the unifying framework of
superstructure Mottness.

\end{abstract}


\maketitle

\section{Introduction}\label{section:intro}

The Mott physics, i.e.\ the transition from band to localized states
due to interaction, plays an important role in  materials containing
strongly correlated electrons. The usual condition for Mott
transition on a lattice requires half-filling, i.e.\ one electron per
site. Due to Coulomb blocking, the electrons are localized and
behave as local moments when the on-site Coulomb repulsion $U$
becomes larger than the bandwidth of the kinetic energy. The
transition metal oxides and rare earths and actinides, involving
narrow bands from the atomic $d$ and $f$ orbitals, are known to
exhibit Mottness -- a remarkable set of electronic properties due to
the proximity to Mott transition~\cite{kotliar,philips}.

Thus far the study of Mott transition and the associated Mottness
has mostly focused on uniform systems. In this paper, we propose and
develop the notion of superstructure Mottness
to describe the Mott physics for inhomogeneous
electronic states. This is important because strongly correlated
electron systems have a propensity towards inhomogeneous phases as a
result of the frustrated kinetic energy. Examples include the
transition metal chalcogenides that form charge density waves
(CDW)~\cite{cdw,cdwstm} and the high-$T_c$ cuprates that exhibit
stripe~\cite{stripes} and checkerboard orders~\cite{checkerboard}.
The superstructure in strongly correlated systems can also come from
structural distortions as in Ca-substituted strontium
ruthenates~\cite{ruthenate} or chemical dopants as in the sodium
cobaltates~\cite{cobaltates}.

Let us consider a commensurate superstructure in two dimensions with
a $p\times q$ supercell. An electron density $n^*=\ell/pq$
corresponds to $\ell$ electrons per supercell. If $\ell$ is an odd
integer, the condition for the superstructure Mott transition is
satisfied by an odd number of electrons per supercell. More
precisely, for $\ell=2k+1$, $k$ of the $M=pq$ subbands are
completely filled, while the $(k+1)$-th subband, hosting one
electron, is at half-filling and can undergo Mott transition. Note
that partial filling of the subbands leading to the semi-metallic
states, which usually happens close to integral filling near the top
and bottom of the subbands for an even number of electrons per
supercell, will not occur here for a generic superlattice potential.
As a result, superstructure Mottness allows the Mott transition to occur far away
from one electron per site. For example, electron correlation is
expected to be weak in the sodium rich phases of Na$_x$CoO$_2$
because of the proximity to a band insulator at $x=1$. To the
contrary, strong correlation effects are observed, possibly due to
the superstructures induced by Na
dopants~\cite{cobaltates,naorder,marianetti,gao} that allow
superstructure Mottness.  Another example is the recent observation of orbital
selective Mott transition in Ca$_{2-x}$Sr$_x$RuO$_4$ near
$x=0.2$~\cite{ruthenate-arpes}; the $d_{xy}$ orbital hosting 1.5
electrons becomes Mott localized due to the $\sqrt{2}\times\sqrt{2}$
superstructure induced by RuO$_6$ octahedral
distortion~\cite{ruthenate} such that there are 3 electrons per
supercell. A third example is that the $\sqrt{13}\times\sqrt{13}$
CDW reconstruction on the surface of 1T-TaSe$_2$ leads to one
electron per supercell and causes a Mott transition at low
temperatures which has been observed recently by scanning tunneling
spectroscopy~\cite{cdwstm}.

In addition to Mott localization, another paradigm in the physics of
strong electronic correlation is Wigner crystallization which, in
contrast, is due to strong long-range Coulomb interaction $V$ that
causes electrons to crystallize (charge order) by Coulomb jamming.
The electron crystallization into a charge ordered Wigner lattice is
another form of electronic superstructures. Wigner localization
happens even classically for low density electrons in the continuum.
What happens to lattice electrons when both $U$ and $V$ are large
and away from half-filling, i.e.\ the Wigner-Mott transition, is
largely unexplored theoretically because of the lack of
nonperturbative many-body methods for treating the finite-range,
inter-site Coulomb repulsion $V$. The basic difficulty is that the
Hilbert space becomes nonlocal in the presence of the inter-site
$V$, which defies the strong-coupling approaches based on the
single-site Hilbert space containing four states representing an
empty site, an up-spin electron, a down-spin electron, and a doubly
occupied site. Previous approaches used a single-site Hilbert space
Gutzwiller projection or dynamical mean field theory (DMFT) to treat the
onsite $U$, but decoupled the inter-site $V$-term using the
weak-coupling Hartree approximation \cite{pietig,mckenzie,camjayi}
which becomes unphysical when $V$ is large since the electrons would
avoid the high self-energy cost by not sitting close together.

It turns out that the notion of superstructure Mottness encompasses the physics
of Wigner-Mott transition in a unified framework. Our basic insight
is to envision the lattice as being covered by clusters or
supercells and think of the latter as artificial "atoms" with a
number of effective orbitals representing the sites contained in
each cluster. This maps the problem to that of a multi-orbital
Mott-Hubbard system where $U$ and $V$ play the role of intra- and
inter-orbital interactions. Charge ordered states on the original
lattice simply correspond to orbital order in the artificial atoms.
This remarkable mapping, although approximate, allows us to treat
the effects of $V$ using the nonperturbative strong coupling
approach for the multi-orbital Hubbard model and thus one is led to
embrace a possible emergent Mott transition with increasing $U$ and
$V$ at the filling fraction of one electron per cluster (supercell).
Such a phenomenon of localization and local moment formation
into clustered superstructures is another form of superstructure Mottness that
unifies the Mott and Wigner physics.

\begin{figure}
    \begin{center}
       \fig{3.3in}{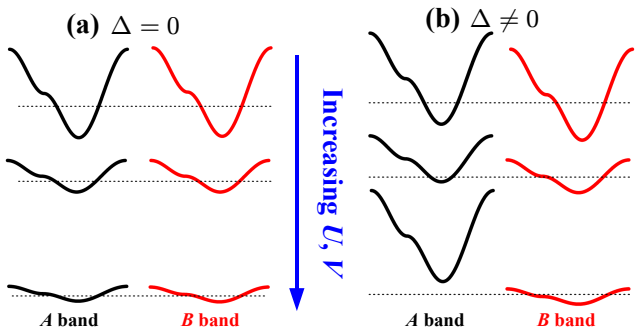}
           \caption{(Color online) (a) and (b) Schematics
                    of band evolution with interaction.}
           \label{fig:schem}
    \end{center}
\end{figure}

The above picture amounts to truncating the nonlocal Hilbert space
as a product of that of the clusters or supercells. The interactions
between the electrons in the neighboring clusters are treated on
equal footing. We extend the well established Gutzwiller projected
wavefunction approach and develop a cluster Gutzwiller method.
Explicit model calculations are performed to test the ideas of
superstructure Mottness. Specifically, we study a minimal one-band, ionic Hubbard
model with intersite correlations, i.e.\ the $t$-$U$-$V$-$\Delta$
model. In addition to the onsite Hubbard $U$ and the nearest
neighbor (NN) Coulomb interaction $V$, an onsite two-sublattice
potential $\Delta_i$ is included for possible external potentials
due to lattice distortion and dopant ions. We map the model to a
generic two-orbital Hubbard model on a superlattice that contains
two sites per supercell. The Hubbard $U$ serves as the intra-orbital
Coulomb $U$; the inter-site $V$ becomes the inter-orbital repulsion
$U^\prime=V$, whereas the superlattice potential $\Delta_i$ turns
into an effective crystal field $\Delta_\alpha$, with $\alpha=A,B$
labeling the two ``atomic'' orbitals with a crystal field splitting
$\Delta=\Delta_A-\Delta_B$. Using the cluster Gutzwiller projection
of the multioccupancy states in a supercell, in analogy to the
multiorbital Gutzwiller projection
method~\cite{multibandgutzwiller}, we study the ground state phase
diagram at quarter filling in the parameter space of $U$, $V$, and
$\Delta$. Fig.~\ref{fig:schem} shows the schematics of two classes
of localization transitions with increasing $U$ and $V$. For
$\Delta=0$, the two bands remain degenerate and undergo the
transition simultaneously from a uniform correlated metal to a Mott
insulator without charge (i.e.\ orbital) ordering. For
$\Delta\neq0$, we find a transition from a CDW metal involving both
bands to a charge ordered insulator, where one of the two bands
undergoes an orbital selective Mott transition~\cite{osmt}. We
elucidate the mechanism by which the superlattice potential enhances
the correlation effects and the tendency towards local moment
formation, and reveal a deeper connection among the strongly
correlated inhomogeneous electronic states, the Wigner-Mott physics,
and the multiorbital Mott physics that can all be united under the
notion of superstructure Mottness.

The rest of the paper is organized as follows. In section II, we
discuss the model and the mapping to an equivalent multiorbital
Hubbard model on the superstructure. In section III, we describe the
virtual cluster Gutzwiller approximation (VCGA) and study the
superstructure Mott transition in the ionic Hubbard model, i.e.\ the
$t$-$U$-$\Delta$ model at quarter-filling, in the absence of
inter-site Coulomb correlation $V$, In section IV, we extend the
VCGA to the case of finite inter-site correlation $V$ and study the
Wigner-Mott transition in the absence of the ionic potential
$\Delta$, i.e.\ the $t$-$U$-$V$ model at quater-filling, using the
ideas of superstructure Mottness. The results will be compared to the
weak-coupling Hartree approximation. The general case of finite
$\Delta$ and $V$ is presented in section V followed by a brief
summary in section VI.

\section{Model and heuristic discussion}

The Hamiltonian of the $t$-$U$-$V$-$\Delta$ model is given by
\begin{equation}
    \hat{H}=-\sum_{i,j}(t_{ij}-\Delta_i\delta_{ij}) c_{i\sigma}^\dagger
c_{j\sigma}+U\sum_i \hat n_{i\up}\hat
n_{i\dn}+V\sum_{\langle i,j\rangle}\hat n_i \hat
n_j,
\label{eq:H}
\end{equation}
where $c_{i\sigma}^\dagger$ creates an electron at site $i$ of spin
$\sigma$ on a square lattice. We consider electron hoppings between
NN $t_{\langle i,j\rangle}=t$ and next NN $t_{\langle\langle
i,j\rangle\rangle}=t^\prime$. Repeated spin indices are summed.
Eq.~(\ref{eq:H}) is an extended version of the ionic Hubbard model
with NN intersite repulsion $V$ and a lattice ionic potential
$\Delta_i$. We focus on the simplest case where $\Delta_i$, when
present, has a superstructure with two-sites per supercell.

Consider the lattice as a complete coverage by a two-sublattice ($A$
and $B$) superstructure. The two sites per supercell (cluster) can
be viewed as two effective orbitals. The hopping terms can be
written in the equivalent form of two bands having dispersions
$\xi_{\alpha\alpha}(k)=-2t^\prime(\cos k_x+\cos k_y)$ with an
inter-band ($\alpha \ne \beta$) hopping $\xi_{\alpha\beta}=-4t\cos
(k_x/2)\cos (k_y/2)$, where $\alpha,\beta=A,B$ label the two
orbitals and $k\in$ the reduced Brillouin zone of the reciprocal
lattice. The Hamiltonian can be rewritten in a suggestive form
\begin{equation}
\begin{split}
    \hat{H}=& \sum_{\alpha,\beta,
k}\xi_{\alpha\beta}(k)c_{k,\alpha\sigma}^\dagger c_{k,\beta\sigma}
+U\sum_{\alpha,I}\hat n_{I,\alpha\up}\hat
n_{I,\alpha\dn}\\
&+ \sum_{\alpha,I}\Delta_\alpha \hat
n_{I,\alpha}
+U^\prime\sum_{I}\hat n_{I,A}\hat n_{I,B}\\
&+V^\prime\sideset{}{'}\sum_{\langle I,J\rangle}\hat{n}_{I,A}\hat
n_{J,B}.
\end{split}
\label{htwoband}
\end{equation}
where the sum over $I$ runs over the supercells on the lattice.
Eq.~(\ref{htwoband}) has the generic form of a two-band Hubbard
model with the onsite intra-orbital Coulomb repulsion $U$ and the
inter-orbital Coulomb repulsion $U^\prime=V$. The superlattice
potential $\Delta_\alpha$ plays the role of a crystal field and
gives rise to a crystal field splitting $\Delta=\Delta_A-\Delta_B$.
The last term with $V^\prime=V$ accounts for the inter-orbital
Coulomb interaction between the neighboring supercells. Note that
the $U$ and $U^\prime$ are free parameters in our theory and not
related by the Hund's rule coupling $J_H$ in contrast to the usual
multiorbital Hubbard model \cite{castenalli}.

\begin{figure}
\begin{center}
\fig{3.3in}{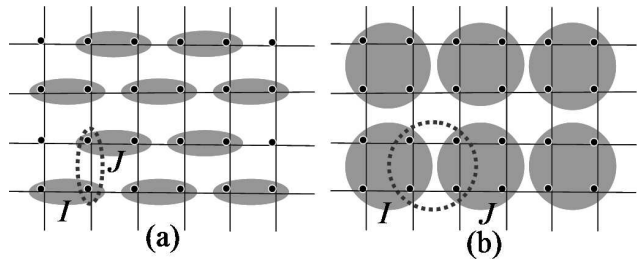}
        \caption{
        Dividing the original lattice into clusters. The dark dots
        represent the original lattice sites while the gray areas
        are the clusters, (a) a two-site supercell and (b) a
        four-site supercell. Here $I$ and $J$ label the position of
        each cluster. The dotted circles represent equivalent
        selection of the supercells.
        }
     \label{fig:clusters}
 \end{center}
\end{figure}

\section{Superstructure Mott transition in the ionic Hubbard model
$\Delta\neq 0$, $V=0$}

We first focus on the ionic Hubbard model in the absence of the
inter-site Coulomb correlation ($V=0$). The correlation effects will
be treated by the Gutzwiller projected wavefunctions. We first
review briefly this variational wavefunction approach to the Hubbard
model and the Gutzwiller approximation as a semi-analytical method
to carry out the projection. The virtual cluster Gutzwiller
approximation will then be introduced to treat the superlattice
modulations induced by the ionic potential $\Delta_i \ne 0$.

\subsection{Gutzwiller approximation to the Hubbard Model}

For the Hubbard model with only onsite repulsion $U$, the Hilbert
space is a direct product of the single-site Fock states: one
empty, two singly occupied (either by an up or a down spin), and
one doubly occupied. The Gutzwiller projected wavefunction
approach is a way of building correlation effects from the
non-interacting wavefunctions by reducing the statistical weight
of double occupation~\cite{Gutzwiller1963}. The trial wavefunction
can be written down formally as $\lvert \Psi \rvert = \hat{P}
\lvert \Psi_0 \rangle$, with $\lvert \Psi_0 \rangle $ being the
non-interacting Slater determinant wavefunction and $\hat{P}$ the
Gutzwiller projection operator. The projection operator $ \hat{P}
= g^{\hat{D}} $, where $\hat{D} = \sum_i \hat{n}_{i\up}
\hat{n}_{i\dn}$ is the double occupation operator of the system
and $g \le 1$ a variational parameter to be determined by
minimizing the ground state energy.  When minimizing the ground
state energy $\langle\Psi\lvert \hat{H} \lvert \Psi \rangle / \langle
\Psi \vert \Psi \rangle$, the average interaction energy is simply
given by $Ud$, with $d$ being the average density of double
occupancy. Since there is a one to one correspondence between $d$
and $g$, one can make $d$ the variational parameter instead of
$g$. The task is then to derive an expression for the kinetic
energy in terms of $d$.  Unfortunately, an analytical expression
for this expectation value is not available other than in one and
infinite dimensions~\cite{Metzner8788}. In two and three
dimensions one has to resort to approximate methods. The
Gutzwiller approximation relates the expectation values in the
projected state with that of the uncorrelated state by a simple
multiplicative factor, the Gutzwiller factor (GF). This is
achieved by ignoring inter-site correlations in evaluating the
expectation values with $\lvert \Psi \rangle $,
\begin{equation}
    \langle \Psi \lvert \sum_{i,j} c_{i\sigma}^\dagger c_{j\sigma}
    \rvert \Psi \rangle / \langle \Psi \vert \Psi \rangle
    \approx g_\sigma^2 \langle \Psi_0 \lvert \sum_{i,j}
    c_{i\sigma}^\dagger c_{j\sigma} \rvert \Psi_0 \rangle,
    \label{projection}
\end{equation}
where $g_\sigma$ is the GF for the kinetic hopping term and
translational invariance of the lattice is assumed. The exact
form of $g_\sigma$ depends on the trial wavefunction $\lvert \Psi_0
\rangle$ and can be obtained by considering how double occupation
affects the hopping process. If $\lvert \Psi_0 \rangle $ is taken as
the Slater determinant state without further symmetry breaking,
$g_\sigma$ takes a simple form in terms of the uniform particle
density and the density of double occupation~\cite{Gutzwiller1965,
Vollhardt1984},
\begin{equation}
    g_\sigma = \left[ \frac{(n_\sigma - d)(1+d-n)}{n_\sigma
    (1-n_\sigma)}\right]^{1/2} + \left[ \frac{d(n_{\bar{\sigma}} -
    d)}{n_\sigma(1-n_\sigma)}\right]^{1/2}.
\end{equation}
As a result, the minimization of the ground state energy for the
original Hubbard Hamiltonian translates to the minimization of the
following renormalized mean field Hamiltonian,
\begin{equation}
    \hat{H}_{\mathrm {GA}} = -\sum_{i,j} g_\sigma^2 t_{ij}
    c_{i\sigma}^\dagger c_{j\sigma} + \sum_{i\sigma}\lambda_\sigma
    (\hat{n}_{i\sigma} - n_\sigma) + NUd,
\end{equation}
where the hopping integrals are renormalized by a factor of
$g_\sigma^2$ from the original ones. Here $N$ is the number of sites
and $\lambda_\sigma$ is a Lagrange multiplier that keeps the
particle density unchanged before and after the projection. Under
the Gutzwiller approximation, the ground state energy of the
original Hubbard model is approximated by that of the mean field
Hamiltonian $\hat{H}_{\mathrm {GA}}$. The correlation effects are
treated nonperturbatively since there are no self-energy corrections
that scale with $U$ in contrast to the weak coupling Hartree-Fock
approach. Since the bandwidth (hopping) is renormalized in
$\hat{H}_{\mathrm {GA}}$, this approach is also known as the
renormalized mean field approximation~\cite{Zhang1988}.

\subsection{The Virtual Cluster Gutzwiller Approximation}

Several authors have extended the original Gutzwiller approximation
to handle lattice translational symmetry breaking using different
schemes~\cite{cli2006,Wang2006,Ko2007,Fukushima2008}. In the
spatially unrestricted Gutzwiller approximation
(SUGA)~\cite{cli2006,Li2009}, the GF for the hopping renormalization
is generalized to depend on the local charge densities and double
occupations at the sites connected by the hopping process.
Specifically, $g_{ij\sigma}$ for the hopping from site $i$ to site
$j$ is a product $g_{i\sigma}\cdot g_{j\sigma}$, with $g_{i\sigma}$
given by
\begin{equation}
    g_{i\sigma}=
     \left[\frac{\left(n_{i\sigma} - d_{i}\right)
               \left(1+d_{i}-n_{i}\right)}
              {n_{i\sigma}\left(1-n_{i\sigma}\right)}
     \right]^{1/2}
        + \left[
                \frac{d_{i}\left(n_{i\bar{\sigma}}
                      - d_{i}\right)}
              {n_{i\sigma}\left(1-n_{i\sigma}\right)}
         \right]^{1/2},
    \label{eq:SUGA}
\end{equation}
where $d_i$ is the double occupation at site $i$, $n_{i\sigma}$ is
the spin dependent density, and $n_i = \sum_\sigma n_{i\sigma}$. The
latter has spatial modulations due to the presence of superlattice
potential $\Delta_i$. The Mott transition in a given superstructure
at appropriate filling fractions can be analyzed using the SUGA and
the GF in Eq.~(\ref{eq:SUGA}).

\begin{figure}
\begin{center}
\fig{3.3in}{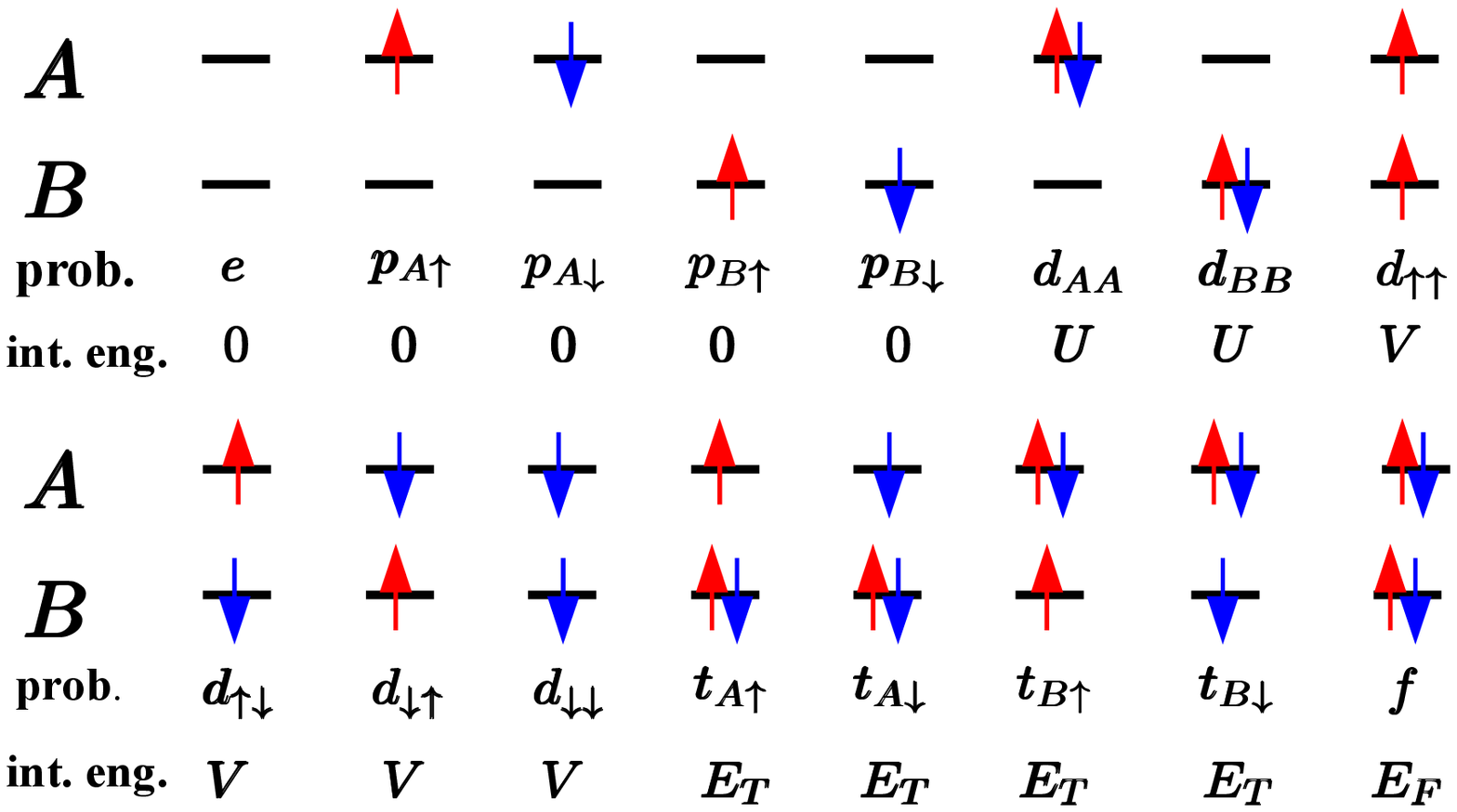}
        \caption{(Color online)  The local Fock space of a
        two-site cluster. The expectation probability of each
        state and the corresponding interaction energy are also
        given. Here $E_T = U+2V$ and $E_F = 2U+4V$.
        }
        \label{fig:2orbital}
   \end{center}
\end{figure}

In order to gain more physical insights into the problem and
establish the notion of superstructure Mottness, we develop a virtual cluster
approach. For a periodic superlattice potential $\Delta_i \ne 0$,
the original lattice is divided into supercells. It is important to
note that there are many equivalent ways to choose the supercell
clusters that cover the whole lattice. A particular two-site
supercell and a four-site supercell coverage of the original lattice
is shown in Figs.~\ref{fig:clusters}(a) and \ref{fig:clusters}(b).
Thus, a lattice site is labeled by $(I,i)$ where the capital letter
$I$ specifies the supercell and $i$ denotes the position inside the
supercell. The Hamiltonian in Eq.~(\ref{eq:H}) can be written as
$\hat{H} = \hat{H}_{\mathrm K} + \hat{H}_{\text{local}}$, where the
``kinetic'' part $\hat{H}_{\mathrm K}$ and the ``local''
interaction
part $\hat{H}_{\text{local}}$ are given by,
\begin{subequations}
    \begin{align}
        \hat{H}_{\mathrm K} =& -\sum_{I,i; J,j}
    t_{ij} c_{Ii\sigma}^\dagger c_{Jj\sigma},
    \label{eq:Hinter}
    \\
    \hat{H}_{\text{local}} =& U\sum_{I,i}\hat{n}_{Ii\up}\hat{n}_{Ii\dn}
    +\sum_{I,i}\Delta_i\hat{n}_{Ii}.
    \label{eq:Hlocal}
    \end{align}
\end{subequations}
In the spirit of Gutzwiller projected wavefunction, the effect of
$\hat{H}_{\text{local}}$ is to reduce multi-occupation in each
supercell. For example, the Fock states of a two-site supercell are
shown in Fig.~\ref{fig:2orbital}, where 11 of the 16 states in a
cluster have more than one particle. The operators for these
multiplets are denoted by ${\hat{\mathbb{M}}}_\gamma$, $\gamma=1,
\dots, \Gamma$, with $\Gamma$ the dimension of the multi-occupancy
space ($\Gamma=11$ in the two-site cluster example). A trial
wavefunction can then be written as $\lvert \Psi \rangle = \hat{P}
\lvert \Psi_0 \rangle$, where the projection operator $\hat{P}$ is a
product of the cluster projection operators, $\hat{P}_I$, i.e.,
$\hat{P}=\prod_I\hat{P}_I$. The cluster projection operator
$\hat{P}_I$ reduces multi-occupation within each cluster $I$ and is
written down as,
\begin{equation}
    \hat{P}_I=\prod_{i\sigma}y_{i\sigma}^{\hat n_{i\sigma}}
              \cdot \prod_{\gamma=1}^\Gamma g_{\gamma}^{\hat{\mathbb{M}}_\gamma},
\label{projector}
\end{equation}
where the supercell index $I$ is omitted on the right hand side
because of translation symmetry on the superlattice. Here the $y$'s
are fugacities that keep charge densities unchanged upon projection
at each site and the  $g_{\gamma}$'s are variationally determined
weighting factors for the projection.

The hopping processes in Eq.~(\ref{eq:Hinter}) is then renormalized
by the projected wave function as in Eq.~(\ref{projection}) and the
GF is calculated by ignoring inter-cell correlations. However, it is
important to note that for a fixed choice of a supercell coverage of
the lattice, $\hat{H}_{\mathrm K}$ includes hopping between
different supercells/clusters ($I\ne J$) as well as hopping within
the same cluster ($I=J$). Conventional cluster approaches, such as
the one used by Lechermann et al., do fix the choice of the
supercells. This, however, leads unavoidably to dimerization since
the inter-cluster hopping and intra-cluster hopping cannot be
treated on equal footing and acquire different renormalization
factors \cite{Lechermann2007}. Here we propose a \emph{virtual
cluster} approach. We do not fix a real space cluster coverage of
the lattice in \emph{a priori}. Instead, the choice of the clusters is
associated with the Hilbert space and arises when evaluating the
expectation values of the hopping terms using the projected
wavefunction in Eq.~(\ref{projection}). Indeed, in order to
implement the Gutzwiller approximation, the hopping term must be
between different clusters (i.e.\ inter-cluster) and the correlation
between the clusters must be ignored. Our insight is that for any
two sites connected by the hopping process on the original lattice,
one can always find a particular choice of supercell coverage such
that the two sites reside in two different supercells. As an
example, choosing the cluster marked by the dotted circles in
Figs.~\ref{fig:clusters}(a) and \ref{fig:clusters}(b) turns the
sites within the cluster (gray) into ones residing in different
supercells.

Specifically, consider the hopping terms in Eq.~(\ref{eq:Hinter}).
In the term where $I=J$, it is always possible to choose an
equivalent cluster coverage such that the hopping is between sites
in different clusters $I^\prime \ne J^\prime$. When evaluating the
expectation value of this hopping term using the projected wave
function, one can simply rearrange the product of the cluster Fock
states and write $\hat{P}=\prod_{I^\prime}\hat{P}_{I^\prime}$ and
arrive at the same result as the case when $I\ne J$. It is in this
sense, we term this approach as the virtual cluster Gutzwiller
approximation (VCGA). One of the most notable advantages of this
approach is that the same Gutzwiller renormalization factor is
obtained for the hopping integral between any two sites regardless
of the choice of the supercell coverage of the original lattice and,
as a result, the difficulty associated with the unphysical
dimerization tendency is removed.

The remaining steps for deriving the GF is the same as that of the
multi-orbital Hubbard model with only on-site density-density
interactions \cite{multibandgutzwiller}. In this sense, one can
regard each supercell as an artificial ``atom'' and the lattice
sites within the cluster as the associated atomic ``orbitals'', and
we will use orbitals and sites within a cluster interchangeably. We
obtain
\begin{equation}
    \begin{split}
    \langle c_{{I}\alpha \sigma}^\dagger
    c_{{J}\beta\sigma}\rangle
    &=\frac{\langle\hat{P}_{{I}} c_{{I}\alpha\sigma}^\dagger
    \hat{P}_{{I}}\hat{P}_{{J}} c_{{J}\beta\sigma}^\dagger
    \hat{P}_{{J}}\rangle_0}{\langle\hat{P}_{{I}}^2
    \hat{P}_{{J}}^2\rangle_0}
    \\
    &= \frac{\langle\hat{\mathbb{W}}_{{I}\alpha\sigma}
    \hat{\mathbb{W}}_{{J}\beta\sigma} \rangle_0}
    {\langle\hat{P}_{{I}}^2
    \hat{P}_{{J}}^2\rangle_0}\cdot \langle c_{{I}\alpha \sigma}^\dagger
    c_{{J}\beta\sigma}\rangle_0
    \\
    &= \frac{\langle\hat{\mathbb{W}}_{{I}\alpha\sigma}
    \rangle_0} {\langle \hat{P}_{{I}}^2\rangle_0}
    \frac{\langle\hat{\mathbb{W}}_{{J}\beta\sigma} \rangle_0}
    {\langle \hat{P}_{{J}}^2\rangle_0} \cdot \langle c_{{I}\alpha
    \sigma}^\dagger
        c_{{J}\beta\sigma}\rangle_0
        \\
    &= g_{{I}\alpha\sigma} g_{{J}\beta\sigma}
      \langle c_{{I}\alpha \sigma}^\dagger
      c_{{J}\beta\sigma}\rangle_0,
   \end{split}
   \label{eq:interchopping}
\end{equation}
where $\langle \cdots \rangle_0$ denotes expectation values in the
unprojected state $\lvert \Psi_0\rangle$ and the equalities hold
only when inter-cluster correlations are ignored in the spirit of
the Gutzwiller approximation. Here
$\hat{\mathbb{W}}_{I\alpha\sigma}$ is defined as,
\begin{equation}
    \hat{\mathbb{W}}_{{I}\alpha\sigma} c_{{I}\alpha\sigma}
    ^\dagger =  \hat{P}_{{I}}
    c_{{I}\alpha\sigma}^\dagger
    \hat{P}_{{I}},
    \label{eq:Wdef}
\end{equation}
and $g_{I\alpha\sigma} = \langle\hat{\mathbb{W}}_{I\alpha\sigma}
\rangle_0/\langle \hat{P}_{I\alpha\sigma}^2 \rangle_0$. Because of
translational symmetry of the superlattice, $g_{I\alpha\sigma}$ is
independent of the supercell index $I$, i.e., $g_{I\alpha\sigma} =
g_{\alpha\sigma}$.

In the rest of this section we focus on a two-site cluster on a
square lattice, compatible with a superlattice potential
\begin{equation} \Delta_i
    = \frac{\Delta}{2}(-1)^{i_x + i_y}. \label{eq:ionic2site}
\end{equation}
The two sites inside a cluster, i.e.\ the two orbitals are labeled
as $A$ and $B$. The $\Delta$ in the above equation has the meaning
of a crystal field splitting between the orbitals. The
multi-occupation projection operators $\mathbb{M}_{\gamma}$'s are as
follows (suppressing the supercell index), $\hat{D}_{\alpha\alpha} =
\hat{D}_{\alpha}\hat{E}_{\bar{\alpha}}$ -- intraorbital doublon
projection operator for the state with a doubly occupied $\alpha$
orbital (created by $\hat{D}_{\alpha}$) and an empty orbital
(created by $\hat{E}_{\bar{\alpha}}$) ;
$\hat{D}_{\sigma\sigma^\prime}=\hat{Q}_{\alpha\sigma}
\hat{Q}_{\bar{\alpha}\sigma'}$ -- interorbital doublon projection
operator for the state with two singly occupied orbitals of spin
projections $\sigma$ (created by $\hat{Q}_{\alpha\sigma}$) and
$\sigma^\prime$ (created by $\hat{Q}_{\bar{\alpha}\sigma'}$);
$\hat{T}_{\alpha\sigma} = \hat{Q}_{\alpha\sigma}
\hat{D}_{\bar{\alpha}}$ -- triplon projection operator for the state
with a singly occupied $\alpha$ orbital with spin $\sigma$ and a
doubly occupied orbital; and $\hat{F} = \hat{D}_{\alpha}
\hat{D}_{\bar{\alpha}}$ -- quadruplon projection operator for the
state with both orbitals doubly occupied.  The projection operators
for singly occupied states are denoted by $\hat{P}_{\alpha\sigma}$
where the $\alpha$ orbital with spin $\sigma$ is occupied while the
other orbital is empty, and projecting to the configuration where
both orbitals are empty is accomplished by $\hat{E} =
\hat{E}_{\alpha} \hat{E}_{\bar{\alpha}}$.  The expectation values of
these operators in the projected state will be denoted by a lower
case letter. For example, $\langle \hat{T}_{\alpha \sigma} \rangle =
t_{\alpha \sigma}$, see Fig.~\ref{fig:2orbital}. The definition of
the single-orbital projection operators are the doubly occupied
$\hat{D}_{\alpha} = \hat{n}_{\alpha\up} \hat{n}_{\alpha\dn}$, singly
occupied $\hat{Q}_{\alpha\sigma} = \hat{n}_{\alpha\sigma} (1 -
\hat{n}_{\alpha\bar{\sigma}})$ and empty orbital $\hat{E}_{\alpha} =
(1-\hat{n}_{\alpha\up}) (1 - \hat{n}_{\alpha\dn})$. The expectation
values of these single-orbital projection operators in the projected
states are also denoted by a lower case letter, e.g., $\langle
\hat{Q}_{\alpha\sigma} \rangle = q_{\alpha\sigma}$. It is seen from
Eq.~(\ref{eq:interchopping}) that the GF, $g_{\alpha\sigma}$,
depends on the fugacities $y_{\alpha\sigma}$ and $g_{\gamma}$. By
expressing $y_{\alpha\sigma}$ and $g_\gamma$ in terms of the
occupation probabilities, we obtain the Gutzwiller factors
\cite{Li2009},
\begin{subequations}
\begin{align}
g_{A\sigma} =&\bigg(\sqrt{ep_{A\sigma}}+\sqrt{d_{AA}p_{A\bar{\sigma}}}
               +\sqrt{d_{\sigma\sigma}p_{B\sigma}}+
               \sqrt{d_{\sigma\bar{\sigma}}p_{B\bar{\sigma}}} \nonumber
               \\
               \phantom{=}&
               +\sqrt{d_{BB}t_{A\sigma}}+\sqrt{d_{\bar{\sigma}\sigma}t_{B\sigma}}
               +\sqrt{d_{\bar{\sigma}\bar{\sigma}}t_{B\bar{\sigma}}}
               +\sqrt{t_{A\bar{\sigma}}f}\bigg) \nonumber
               \\
               \phantom{=}&
               \big{/}\sqrt{n_{A\sigma}(1-n_{A\sigma})},
               \label{ga}\\
g_{B\sigma} =&\bigg(\sqrt{ep_{B\sigma}}+\sqrt{d_{BB}p_{B\bar{\sigma}}}
               +\sqrt{d_{\sigma\sigma}p_{A\sigma}}
               +\sqrt{d_{\bar{\sigma}\sigma}p_{A\bar{\sigma}}} \nonumber
               \\
               \phantom{=}&
               +\sqrt{d_{AA}t_{B\sigma}}
               +\sqrt{d_{\sigma\bar{\sigma}}t_{A\sigma}}
               +\sqrt{d_{\bar{\sigma}\bar{\sigma}}t_{A\bar{\sigma}}}
               +\sqrt{t_{B\bar{\sigma}}f}\bigg)\nonumber
               \\
               \phantom{=}&
               \big{/}\sqrt{n_{B\sigma}(1-n_{B\sigma})}.
               \label{gb}
\end{align}
\end{subequations}
In addition, the occupation probabilities satisfy the following
constraints,
\begin{subequations}
    \begin{align}
&p_{A\sigma}  =
n_{A\sigma}-d_{AA}-\sum_{\sigma^\prime}d_{\sigma\sigma^\prime}
                 -t_{A\sigma}-\sum_{\sigma^\prime}
                 t_{B\sigma^\prime}-f,  \label{eq:pasigma} \\
&p_{B\sigma}  =
n_{B\sigma}-d_{BB}-\sum_{\sigma^\prime}d_{\sigma^\prime\sigma}
                 -t_{B\sigma}-\sum_{\sigma^\prime}
                 t_{A\sigma^\prime}-f,  \label{eq:pbsigma} \\
&e  =  1-\sum_\alpha n_{\alpha}+\sum_\alpha
d_{\alpha\alpha}+\sum_{\sigma\sigma^\prime}d_{\sigma\sigma^\prime}
    +2\sum_{\alpha,\sigma}t_{\alpha\sigma}+3f,
    \label{eq:count}
\end{align}
\end{subequations}
where Eqs.~(\ref{eq:pasigma}) and (\ref{eq:pbsigma}) are just
alternative expressions of the electron densities in terms of the
occupation probabilities and Eq.~(\ref{eq:count}) is the
completeness of the \emph{cluster} Fock states. These results are
consistent with those obtained for the multiorbital Hubbard
model~\cite{multibandgutzwiller}.

\begin{figure}
\begin{center}
\fig{3.3in}{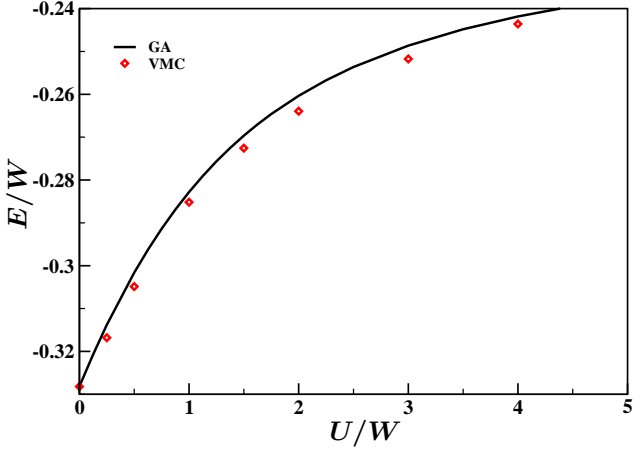}
        \caption{(Color online) Comparison of ground state energy
        of the Hubbard model at quarter filling treated with
        Gutzwiller approximation and variational Monte
        Carlo~\cite{shiba}. $V=\Delta=0$, $W=4t$.}
        \label{fig:engcmpVMC}
   \end{center}
\end{figure}

As an important check for the validity of the VCGA, we verified that
the expressions in Eqs.~(\ref{ga}) and (\ref{gb}) are equivalent to
the results from the SUGA when inter-site correlations are ignored.
To show this, one can make use of the fact that in the absence of
inter-site correlations the multi-occupation probabilities in the
virtual cluster approach are simple products of the occupation
probabilities in SUGA. For example, the above defined
triple-occupation $t_{A\sigma}$ is the product of the double
occupation $d_B$ and single occupation $q_{A\sigma}$, i.e.\,
$t_{A\sigma} = d_B\cdot q_{A\sigma}$.  By substituting all the
multi-occupation probabilities in Eqs.~(\ref{ga} and \ref{gb}) one
recovers the results given in Eq.~(\ref{eq:SUGA}).

\subsection{Superstructure Mott Transition in the Ionic Hubbard Model}

We next present the results for the ionic Hubbard model obtained
with the VCGA at quarter filling for the superlattice potential of
Eq.~(\ref{eq:ionic2site}). Since the system is far away from half
filling, no magnetic instability is present and we will focus on the
paramagnetic solutions. The effective Hamiltonian in the VCGA
discussed above is given by,
\begin{align}
&\hat{H}_{\mathrm {GA}} =
\sum_{k,\alpha\beta}g_{\alpha\sigma}g_{\beta\sigma}
\xi_{\alpha\beta}(k)c_{k,\alpha\sigma}^\dagger
c_{k,\beta\sigma} +\sum_{k,\alpha}\Delta_\alpha\hat{n}_{k,\alpha}
\nonumber
\\
&+N_cU(\sum_{\alpha}d_{\alpha\alpha}
+\sum_{\alpha\sigma}t_{\alpha\sigma}+2f)
+\sum_{k,\alpha\sigma}\lambda_{\alpha\sigma}
(\hat{n}_{k,\alpha\sigma}-n_{\alpha\sigma}), \label{eq:HGADelta}
\end{align}
where $N_{\mathrm c}$ is the number of clusters and
$\lambda_{\alpha\sigma}$'s are the Lagrange multipliers that keep the
particle density in each orbital unchanged following the projection.
They originate from and have a one to one correspondence to the fugacities
in the projection operator in Eq.~(\ref{projector}). The ground
state properties are determined by minimizing $\langle \hat{H}_{\rm
{GA}}\rangle $ and solving for
$(e,p_{\alpha\sigma},d_{\alpha\alpha},d_{\sigma\sigma^\prime},
t_{\alpha\sigma},f)$ and $\lambda_{\alpha\sigma}$ self-consistently.

To test the algorithm, we first compute the ground state energy per
site at quarter-filling as a function of $U/W$, where $W=4t$ is the
half-bandwidth, for the Hubbard model (i.e.\ set $\Delta=0$).
Fig.~\ref{fig:engcmpVMC} shows that the later agrees remarkably well
with the results obtained using the variational Monte Carlo method
to carry out the projection~\cite{shiba}. From the perspective of
superstructure Mottness, it is clearly seen from Eq.~(\ref{eq:HGADelta}) that the
$U$ term alone does not reduce the interorbital double occupancy
$d_{\sigma\sigma^\prime}$. As a consequence, $U$ alone cannot drive
a Mott transition away from half filling since the Gutzwiller
factors in Eqs.~(\ref{ga}) and (\ref{gb}) are always finite and the
system is in a correlated metallic phase.

\begin{figure}
    \begin{center}
        \fig{3.3in}{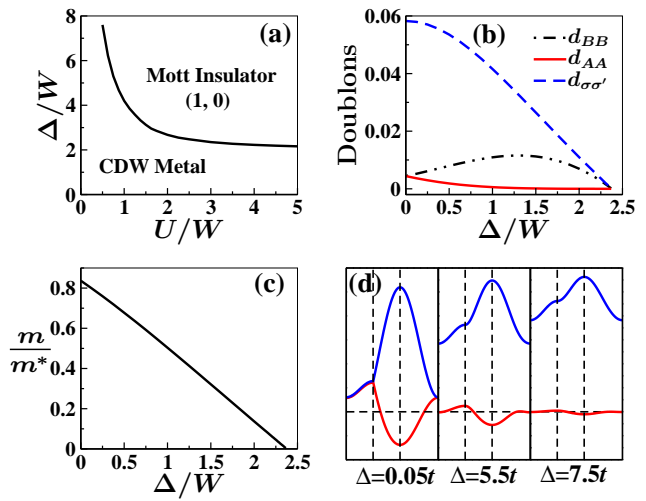}
        \caption{(Color online) (a) Phase diagram on the
        $U$-$\Delta$ plane for $V=0$ and $t^\prime=0$. Doublon
        densities (b) and inverse effective mass (c) as function
        of $\Delta$ for fixed $U=3W$. (d) Band dispersion for
        different $\Delta$, $t' = -0.2t$. $k=(\pi,0)\to (\pi/2,
        \pi/2) \to (0,0) \to (\pi, 0)$.}  \label{fig:UDelta}
    \end{center}
\end{figure}

Switching on $\Delta\ne0$, we determined the phase diagram of the
ionic Hubbard model shown in Fig.~\ref{fig:UDelta}(a) on the
$U$-$\Delta$ plane where a CDW metal at small $(U,\Delta)$ and a
charge ordered insulator for large $(U,\Delta)$ are separated by a
continuous superstructure Mott metal-insulator transition. The
notion of superstructure Mottness offers a physical explanation of the phase
structure through the analogy to multi-orbital Mott-Hubbard systems.
A finite crystal field splitting $\Delta$ induces orbital order
$n_A\ne n_B$, which corresponds to charge ordering in the ionic
Hubbard model. However, below a critical strength of $\Delta$, as
seen in Fig.~\ref{fig:UDelta}(a), the system remains in a metallic
charge density wave (CDW) state no matter how strong $U$ is. This
can be understood since the two bands derived from the two orbitals
are not separated in this case and the inter-orbital hopping is not
suppressed due to the absence of an inter-orbital repulsion
$U^\prime$, which is related to the inter-site correlation $V$ as
shown in Eq.~(\ref{htwoband}).

A sufficiently large crystal field splitting $\Delta$ drives the
system from a CDW metal to a charge ordered insulating state for
large enough $U$. This happens because the two bands gradually
separate with increasing orbital polarization $\delta n=n_B-n_A$.
The lower band becomes half-filled since the system is at
quarter-filling. A large enough $U$ drives a Mott transition in the
half-filled lower quasiparticle (QP) band. Indeed, as $n_A \to 0$
and $p_{A\sigma}\to 0$, Eq.~(\ref{eq:pasigma}) implies that all the
double occupancies approach zero. This leads to $e \to 0$ at quarter
filling from Eq.~(\ref{eq:count}). On the other hand, $n_B \ne 0$
such that from Eq.~(\ref{gb}) the Gutziwller factor $g_{B\sigma} \to
0$ resulting in an orbital selective Mott transition shown in
Fig.~\ref{fig:UDelta}(d). In
Figs.~\ref{fig:UDelta}(b),~\ref{fig:UDelta}(c), the doublon
densities and the inverse effective mass of the lower QP band
$m/m^*$, which equals the QP spectral weight $Z$, are plotted as a
function of $\Delta$ at a fixed large $U=3W$. The continuous
transition at $\Delta_c=2.4W$ is a Brinkman-Rice type
transition~\cite{brinkmanrice}. The vanishing of the coherent QP
spectral weight and the divergence of the effective mass (case shown
in Fig.~\ref{fig:schem}(b)) on approaching the transition from the
metallic side are clear signatures of Mottness.

The curvature of the phase boundary at large ionic potential
$\Delta$ can also be understood. In this regime, the crystal field
splitting is much larger than the hopping $t$ and the bands are well
separated. A simple calculation shows that the bandwidth $w$ of the
lower QP band decreases with increasing $\Delta$, $w\sim
t^2/\Delta$. As a result, the critical $U_c$ for the Mott
transition, which depends on the ratio of $U/w$, decreases as can be
seen in the phase diagram Fig.~\ref{fig:UDelta}(a). The evolution of
the QP band dispersion as a function of $\Delta$ is shown in
Fig.~\ref{fig:UDelta}(d) at fixed large $U=3W$. It clearly
demonstrates that the Mott transition in the ionic Hubbard model
belongs to class (b) of the superstructure Mottness depicted in
Fig.~\ref{fig:schem}.

While the NN hopping $t$ translates into interband hopping in the
two-band model, the NNN hopping $t^\prime$ describes intraband
hopping. We find that including a $t^\prime=-0.2t$ moves the phase
boundary toward smaller $(\Delta,U)$, as can be seen in
Fig.~\ref{fig:UDelta}(d) where a continuous transition takes place
at a smaller $\Delta_c=1.9W$.

\section{Superstructure Mottness in the $t$-$U$-$V$ Model} \label{sec:UV}

In this section, we study the superstructure Mott transition due to
the inter-site Coulomb interaction $V$ in the extended Hubbard, or
the $t$-$U$-$V$ model. To this end, the superlattice potential
$\Delta$ is set to zero. Superstructure Mottness in the presence of both $V$
\emph{and} $\Delta$ will be studied in the next section. The basic
difficulty introduced by the inter-site interaction V is that the
Hilbert space becomes nonlocal, making it inaccessible to the
conventional Gutzwiller projected wavefunction. We show that the
VCGA offers a natural nonperturbative treatment of the Wigner-Mott
transition within the framework of superstructure Mottness.

\subsection{Virtual cluster Gutzwiller approximation for the
$V$-term}

In the Gutzwiller approach, one would like to treat the interactions
as a projection that reduces multi-occupation from an unprojected
Slater-determinate state, i.e.\ $\lvert \Psi \rangle = \hat{P} \lvert
\Psi_0 \rangle$. However, the projection operator $\hat{P}$ cannot
be written down in closed form, since the Hilbert space is
infinitely connected by the inter-site $V$ in Eq.~(\ref{eq:H}). Our
strategy is to truncate the nonlocal Hilbert space as a product of
that of the supercell clusters such that the $U$ and $V$ within a
cluster serve as the intra-orbital $U$ and inter-orbital $U^\prime$
just as in a multi-orbital Hubbard model, which can be treated
nonperturbatively. The projection operator $\hat{P}$ can thus be
written approximately as the product of the cluster projection
operators, $\hat{P}_I$, i.e., $\hat{P}=\prod_I\hat{P}_I$. This
algorithm can be accomplished by the VCGA for the ionic Hubbard
model discussed in section III.A, with a few extensions.

\begin{figure}
\begin{center}
\fig{3.3in}{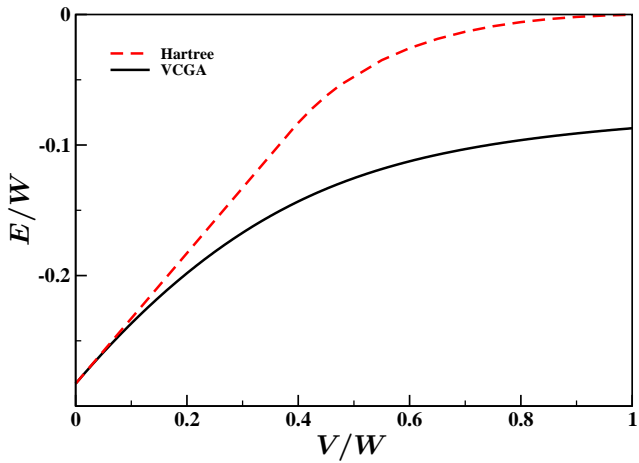}
        \caption{(Color online)  Comparison of the ground state energy
        obtained by VCGA and Hartree approximation of $V$-term at $U=W$.}
        \label{fig:engcompHtree}
   \end{center}
\end{figure}

Although we do not fix a specific supercell structure in the virtual
cluster approach, it is useful to write down the Hamiltonian for a
given coverage of the lattice. Due to the non-local nature of the
$V$-term, the interaction part of the Hamiltonian can no longer be
written as a direct summation over independent supercells. Let us
write the Hamiltonian as the sum of two parts, similar to
Eqs.~(\ref{eq:Hinter}) and (\ref{eq:Hlocal}),
\begin{subequations}
    \begin{align}
        \hat{H}_{\mathrm K} =& -\sum_{I,i; J,j;\sigma}
        t_{ij} c_{Ii\sigma}^\dagger c_{Jj\sigma}
    \label{eq:HinterV}
    \\
    \hat{H}_{\text{inter}} =& U\sum_{I,i}\hat{n}_{Ii\up}\hat{n}_{Ii\dn}
    + V\sum_{I,i;J,j} \hat{n}_{Ii} \hat{n}_{Jj}.
    \label{eq:HlocalV}
    \end{align}
\end{subequations}
The second term in Eq.~(\ref{eq:HlocalV}) contains both the
intra-cluster as well as the inter-cluster $V$-interactions.
Consider now the projection of the hopping term. In the VCGA
discussed before, for a given hopping term between two sites, one
can always choose a cluster coverage such that the two sites reside
in two different clusters. The expectation value of the hopping term
between the projected states can be evaluated according to
Eq.~(\ref{eq:interchopping}). Moreover, to be consistent with the
Gutzwiller approximation, the inter-cluster correlation should be
ignored, which means that the inter-cluster $V$ in
Eq.~(\ref{eq:HlocalV}) should be turned off accordingly. The form of
the cluster projection operator is therefore formally identical to
that in Eq.~(\ref{projector}) for the ionic Hubbard model, since the
intra-cluster $V$-term in Eq.~(\ref{eq:HlocalV}) only leads to
different energies of the multiplets in the cluster Hilbert space.
An explicit example of the $V$-dependent cluster Fock state energies
is given in Fig.~\ref{fig:2orbital} for the case of a two-site
cluster. Consequently, the Gutzwiller hopping renormalization
factors for all hopping amplitudes have the same form and are given
by Eqs.~(\ref{ga}) and (\ref{gb}) in terms of occupation
probabilities for two-site clusters. The inter-cluster $V$-term will
take the same value since the bond becomes intra-cell under a
different choice of the virtual cluster configuration. In short, the
advantage of the virtual cluster formulation is that all bonds
connected by hopping and the inter-site Coulomb interaction V are
treated on equal footing and remain equivalent after Gutzwiller
projection without artificially breaking the lattice translation
symmetry.

\subsection{Wigner-Mott transition as superstructure Mottness}

We now present our results obtained for the extended Hubbard
$t$-$U$-$V$ model at quarter-filling using the VCGA with two-site
clusters. Following the above discussion, the renormalized
mean-field Hamiltonian in the VCGA is given by,
\begin{align}
\hat{H}_{\mathrm {GA}} =&
\sum_{k,\alpha\beta}g_{\alpha\sigma}g_{\beta\sigma}
\xi_{\alpha\beta}(k)c_{k,\alpha\sigma}^\dagger
c_{k,\beta\sigma}
\nonumber \\
&+N_cU(\sum_{\alpha}d_{\alpha\alpha}
+\sum_{\alpha\sigma}t_{\alpha\sigma}+2f)\nonumber \\
&+4N_cV(\sum_{\sigma\sigma^\prime}d_{\sigma\sigma^\prime}
+2\sum_{\alpha\sigma}t_{\alpha\sigma}+4f) \nonumber
\\
&+\sum_{k,\alpha\sigma}\lambda_{\alpha\sigma}
(\hat{n}_{k,\alpha\sigma}-n_{\alpha\sigma}), \label{eq:HGAV}
\end{align}
where the Gutzwiller factors are given by Eqs.~(\ref{ga}) and
(\ref{gb}) with the occupation probabilities satisfying the fermion
counting and completeness equations
(\ref{eq:pasigma}-\ref{eq:count}). The paramagnetic ground state is
obtained by minimize $\langle \hat{H}_{\mathrm GA} \rangle$ with
respect to the variational parameters just as in the ionic Hubbard
model studied in Section III.

\subsubsection{Comparison to Hartree approximation}

We first demonstrate that our strong coupling, nonperturbative
VCGA treatment of the inter-site correlation $V$ is fundamentally
superior to the weak-coupling Hartree approximation of the $V$-term
with a single-site Gutzwiller projection of the on-site repulsion
$U$. To this end, we compare the ground state energy of the
quarter-filled $t$-$U$-$V$ model at $U=W$.
Fig.~\ref{fig:engcompHtree} shows that while the ground state energy
obtained by VCGA is consistent with the weak coupling Hartree result
for small values of $V$, which is in fact a nontrivial check for the
validity of the VCGA, it is significantly lower for all values of
$V$ larger than about 10\% of the bandwidth. The ground state energy
approaches zero for $V\sim W$ in the Hartree approach, signifying a
transition to the charge ordered insulating state. The result of the
VCGA clearly shows that the Hartree approximation grossly
overestimated the charge ordering tendency. This is because the
weak-coupling Hartree-decoupling gives rise to a self-energy for the
fermions that scales with $V$ and becomes unphysical when $V$ is
large. On the other hand, in the nonperturbative VCGA, there is no
self-energy cost that scales with $V$. Thus the uniform metallic
ground state turns out to be much more robust against inter-site
correlations.

While it is clear that the results of the VCGA can be systematically
improved with increasing cluster size to  treat the inter-site
correlations even better, we show below that it already captures the
essence of the superstructure Mott metal-insulator transition with
two-site supercells.

\begin{figure}
\begin{center}\fig{3.3in}{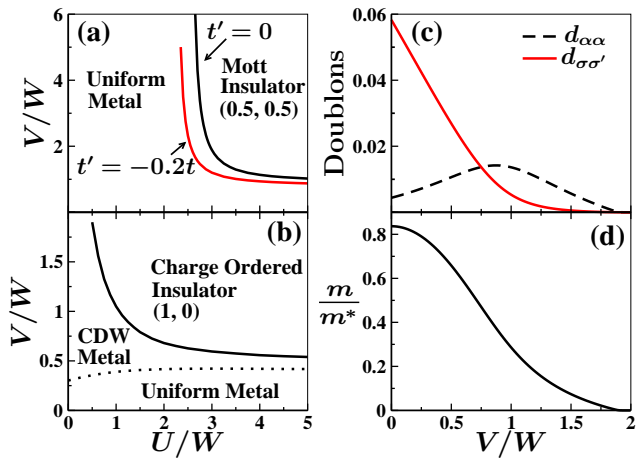}
        \caption{(Color online) (a) Phase diagram of the $t$-$U$-$V$
        model on the $U$-$V$ plane. (b) Phase diagram in the $V$-Hartree
        approximation. Doublon densities (c) and inverse effective mass (d)
        as functions of $V$ at fixed $U=3W$.}
        \label{fig:UV}
        \end{center}
\end{figure}

\subsubsection{Wigner-Mott transition in the $t$-$U$-$V$ model}

The phase diagram obtained using the VCGA at quarter-filling is
shown in Fig.~\ref{fig:UV}(a) on the $U$-$V$ plane. It contains a
uniform strongly correlated metal at small $(U,V)$ and a Mott
insulator at large $(U,V)$. Since the crystal field splitting is
absent, the two degenerate QP bands narrow and undergo a continuous
Mott transition simultaneously upon approaching the phase boundary
from the metallic side, corresponding to the first case of the
superstructure Mottness transition shown in Fig.~\ref{fig:schem}(a).
Figs.~\ref{fig:UV}(c) and \ref{fig:UV}(d) show that double
occupation, the (inverse) QP mass renormalization $m/m^*$ and the QP
coherence weight decrease continuously with increasing $V$, which
are clear signatures of Mottness, and vanish at the critical
$V_c=1.9W$ for $U=3W$. The strongly correlated metallic phase is
stable against the two-sublattice antiferromagnetic (AF) spin
density wave order because the quarter-filled Fermi surface is away
from the AF zone boundary. It is remarkable that for degenerate
bands, the Mott transition driven by $V$ takes place with uniform
charge density without forming a Wigner lattice, suggesting that
superstructure Mottness is a more suitable description than a Wigner crystal.
Since the Mott insulating state has the same energy as the $(0,1)$
charge ordered classical Wigner crystal, an arbitrarily small
crystal field (i.e.\ superlattice potential) would trigger a first
order transition to an orbital ordered state, corresponding to a
charge ordered state in the original model. Interestingly, if the
two bands have different bandwidths, corresponding to anisotropic
NNN hopping $t_A^\prime\neq t_B^\prime$, the Mott insulating state
emerges at large $U$ and $V$ continuously from a CDW metal with a
finite density polarization $\delta n$.

It is useful to further compare these results to  previous studies
where the $V$-term is decoupled by Hartree
approximation~\cite{pietig,mckenzie,camjayi}. The phase diagram in
the latter case obtained by the single-site Gutzwiller approximation
is shown in Fig.~\ref{fig:UV}(b), where a uniform metal at small $V$
transforms into a CDW metal and eventually to a charge ordered
insulator at large $V$.  The topology of this phase diagram is
consistent with that obtained by the dynamical mean field theory at
finite temperatures with a semi-circular density of
states~\cite{camjayi}. However, the CDW metal phase is absent in our
VCGA approach that treats $U$ and $V$ on equal footing. It is likely
that in the Hartree approximation, the fermion self-energy scaling
with $V$ overemphasizes the symmetry breaking CDW ordering tendency
at large $V$ as discussed above.

\begin{figure} \begin{center}\fig{3.3in}{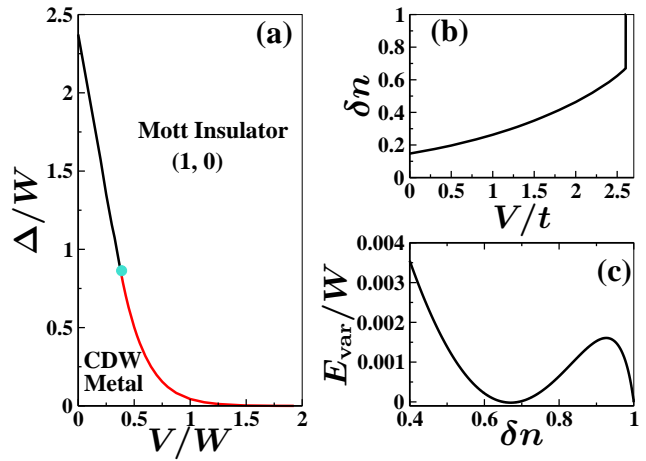}
        \caption{(Color online) (a) Phase Diagram of the $t$-$U$-$V$-$\Delta$
        model on the $V$-$\Delta$ plane at fixed $U=3W$. The circle on the
        boundary separates discontinuous and continuous
        transitions. (b) Charge difference $\delta n= n_B -n_A$ as
        a function of $V$ with $\Delta=t$ and $U=3W$. (c) The
        variational energy as a function of $\delta n$ at $\Delta
        = t$, $V = 2.6t$, and $U=3W$.}  \label{fig:UVDelta}
    \end{center}
\end{figure}

\section{Superstructure Mottness with combined $\Delta$ and $V$} \label{sec:DeltaUV}

Finally, we turn to the most general case of the
$t$-$U$-$V$-$\Delta$ model and study the interplay between the
crystal field splitting $\Delta$ and the extended Coulomb
interaction $V$. Treating the $V$-term using the VCGA in the same
way as in Sec.~\ref{sec:UV} and putting the $\Delta$-term in the
local part of the Hamiltonian in Eq.~(\ref{eq:HlocalV}), we arrive
at the identical form of the Gutzwiller renormalization factors. The
only difference is that the crystal field term should be added in
the renormalized mean-field Hamiltonian in Eq.~(\ref{eq:HGAV}).
Fig.~\ref{fig:UVDelta}(a) displays the phase diagram at $U=3W$ for
the quarter-filled $t$-$U$-$V$-$\Delta$ model. In contrast to the
$V=0$ case shown in Fig.~\ref{fig:UDelta}(a), the coherent motion in
the strongly correlated metallic state is further suppressed by the
inter-site correlation $V$ such that a small (infinitesimal)
$\Delta$ can drive the system to the charge ordered insulating
state, provided $V$ is large. Similarly, the superlattice potential
$\Delta$ is seen to enhance the correlation effects such that the
localization and local moment formation can be induced by a moderate
strength of the inter-site Coulomb interaction $V$.

The phase boundary between the CDW metal and the charge ordered
insulator in Fig.~\ref{fig:UVDelta}(a) changes from a discontinuous
Wigner-Mott transition at small $\Delta$ to a continuous
Brinkman-Rice transition above a critical $\Delta$. The existence of
two types of metal-insulator transition when both $V$ and $\Delta$
are present is consistent with a recent DMFT study of a two-band
Hubbard model with $U=U^\prime$ and $\Delta\neq0$~\cite{antonie}.
Fig.~\ref{fig:UVDelta}(b) shows the charge density difference as a
function of $V$ at $\Delta=t$, displaying a discontinuous jump at a
critical $V_c=2.6t$. The discontinuous transition is a consequence
of degenerate minima in the ground state energy for fixed
$(V,\Delta)$. In Fig.~\ref{fig:UVDelta}(c), we plot the variational
ground state energy $E_{\rm var}$ as a function of the density
difference $\delta n$ at $V=2.6t$ and $\Delta=t$ close to the phase
boundary. The minimum at $\delta n=0.67$ corresponding to a CDW
metal is nearly degenerate to that of the charge ordered insulator
at $\delta n=1$.
Increasing $V$ and $\Delta$ further will make the energy of the
insulating state the global minimum and trigger the first order
transition. The existence of two types of localization transition
into the local moment phase is an important and unique property
associated with superstructure Mottness.

\section{Conclusions}\label{conclude}

We have shown that the superlattice potential in an inhomogeneous
electronic state and the inter-site Coulomb repulsion increase the
correlation effect and the tendency toward localization and local
moment formation. Mapping to the generic multiband Hubbard model
revealed a deeper connection among the strongly correlated
inhomogeneous electronic states, the Wigner-Mott physics, and the
multiorbital Mott-Hubbard physics under the unified notion of
superstructure Mottness. An unbiased treatment of both the on-site and
inter-site Coulomb interactions, e.g. the virtual cluster
approximation developed here within the framework of the Gutzwiller
approximation, is essential as demonstrated for the quarter filled
Hubbard model. We expect that doping away from such unconventional
insulating states will further reveal behaviors of superstructure Mottness,
such as the coexistence of local moment and itinerant carriers. At
dilute electron densities, the competition between itinerant
ferromagnetism and local moment formation may also arise due to
superstructure Mottness in inhomogeneous electron systems.

We thank Hong Ding and Xi Dai for useful discussions. ZW thanks the
Institute of Physics, Chinese Academy of Sciences for hospitality.
This work was supported in part by DOE grant DE-FG02-99ER45747 and
NSF grant DMR-0704545.

\bibliographystyle{apsrev}

\end{document}